\documentclass[aps,prl,showpacs,twocolumn,superscriptaddress]{revtex4}
\usepackage{graphicx}
\usepackage{bm}
\usepackage{amsmath}

\def\be{\begin{equation}}       \def\ee{\end{equation}}
\def\bea{\begin{eqnarray}}      \def\eea{\end{eqnarray}}

\begin{document}
\title{Spin current and magneto-electric effect in non-collinear magnets}
\author{Hosho Katsura}
\email[Electronic address: ]{katsura@appi.t.u-tokyo.ac.jp}
\affiliation{Department of Applied Physics, University of Tokyo,
Hongo, Bunkyo-ku, Tokyo 113-8656, Japan}
\author{Naoto Nagaosa}
\affiliation{Department of Applied Physics, University of Tokyo,
Hongo, Bunkyo-ku, Tokyo 113-8656, Japan} \affiliation{CERC, AIST
Tsukuba Central 4, Tsukuba 305-8562, Japan} \affiliation{CREST,
Japan Science and Technology Agency (JST)}
\author{Alexander V. Balatsky}
\email[Electronic address: ]{avb@lanl.gov} \affiliation{Theoretical
Division, Los Alamos National  Laboratory, Los Alamos, New Mexico
87545 USA}

\begin{abstract}
A new microscopic mechanism of the magneto-electric (ME) effect based on the
spin supercurrent is theoretically presented for non-collinear
magnets. The close analogy between the superconductors (charge
current) and magnets (spin current) is drawn to derive the
distribution of the spin supercurrent and the resultant electric
polarization. Application to the spiral spin structure
is discussed.
\end{abstract}

\maketitle
     The interplay between the magnetism and ferroelectricity is an old
issue since the first prediction of the magnetoelectric (ME) effect
by Curie\cite{Curie}. Later the phenomenological theory of ME effect
was developed by Landau\cite{Landau} and
Dzyaloshinskii\cite{Dzyaloshinskii}. There the symmetry
consideration is essential to classify the all possible ME tensors
depending on the magnetic point group. Especially the time-reversal
(T) and spatial inversion (I) are the key symmetries to the ME
effect. For example, the linear ME effect corresponding to the term
$\alpha_{ij} M_i P_j$ (${\vec P}$:polarization, ${\vec
M}$:magnetization) is allowed only when both $T-$ and $I-$symmetries
are broken. However, on the other hand, the microscopic quantum
theory of ME effect has not yet been fully  developed although
several scenarios for particular materials such as Cr$_2$O$_3$ have
been proposed\cite{Date,Rado}. For this particular material, the
change of the anisotropy energy, exchange, and $g$-value due to the
electric field have been proposed for the origin of the parallel ME
effect. The transverse ME effect, on the other hand, can not be
explained by these mechanisms. For a spiral spin magnet such as
ZnCr$_2$Se$_4$, the electric field induced Dzyaloshinskii-Moriya
(DM) interaction \cite{DM1,DM2} has been proposed for the transverse
ME effect \cite{shiratori}. Another well-studied ME material is
(Ga,Fe)O$_3$, where the ME coefficient is an order of magnitude
larger than in Cr$_2$O$_3$, which appears mostly in the
anti-symmetric transverse channel.  This ME effect was  attributed to
the presence of toroidal moment $ {\vec T} = { {\mu_B} \over 2}
\sum_a {\vec r}_a \times {\vec S}_a$\hspace*{5pt}[9].
\\ \hspace*{8pt}We will argue that ME effect and spin current are directly related. The spin current ${\vec j}_s$ has attracted
revived interests recently in the context of spintronics in
semiconductors. In contrast to the charge current, it is $T$-even
since the spin polarization is also reversed together with the
direction. Therefore from the viewpoint of pure symmetry, ${\vec
j}_s$ belongs to the same class as the electric polarization ${\vec
P}$, and it is natural to expect the coupling between these two. In
fact, the electric field induced dissipationless spin current driven
by the Berry phase curvature has been proposed for the
semiconductors  \cite{spintronics}.
In magnets, the spin current is analogous to the superfluid current,
i.e., spin supercurrent, associated with the spin rigidity
\cite{supercurrent}. In this paper, we present a new mechanism for
the ME effect due to the spin supercurrent with the non-collinear
spin structure such as the spiral state\cite{Yoshimori,Nagamiya}.
The idea is that the spin supercurrent is induced between the two
spins with generic non-parallel configurations, which induces the
electric polarization. This is the dual effect to DM
interaction\cite{DM1,DM2} and/or Aharonov-Casher effect\cite{AC}.
The vector potential coupled to the spin current is the vector
product of electric field/electric polarization and the direction of
the bond connecting the two spins\cite{Loss}. This allows one to
develop  a microscopic theory of the electric field induced  DM
interaction \cite{shiratori}.  Applications to the spiral magnetic
structure are also  discussed.

    Let us first review the analogy of the magnetically ordered state to
the superconducting state. The key to understand the superconductivity is
the canonical conjugate relation between the charge $n$ and the Josephson
phase $\varphi$, i.e.,
$
[ n_i, \varphi_j ] = i\delta_{ij},
$
where $i,j$ are the indices for the site. Similar relation exists in the
quantum spin operators as
$
[S^z_i, \theta_j] = i \delta_{ij},
$
where $S^z$ is the $z$-component of the spin operator while $\theta$ is the
angle of the vector $(S^x,S^y)$ measured from the $x$-axis. This makes the
link between XY spin model and that of superconductivity.
The Hamiltonian for XY model reads

\begin{equation}
H_{XY} = - \sum_{<ij>}
{{J_{\perp ij} } \over 2} ( S^+_i S^-_j + S^-_i S^+_j)
\end{equation}
and the spin supercurrent $j^s_{ij}$ defined so as to satisfy
$\partial
S_i^z/ \partial t  = ( 1 / i \hbar) [ S_i^z, H]
= - \sum_j j^s_{j i}$ is given by
\begin{equation}
j^s_{ij} = iJ_{\perp ij }  ( S^+_i S^-_j - S^-_i S^+_j)
\end{equation}
Putting $(S^x_j, S^y_j) = S ( \cos \theta_j, \sin \theta_j)$,
We obtain
\begin{equation}
j^s_{ij} = J_{\perp ij}S^2 \sin( \theta_i - \theta_j)
\end{equation}
where $J_{\perp}S^2$ corresponds to the spin stiffness, i.e., rigidity.
Equation (3) is analogous to the Josephson equation.

To go further with this analogy, the next question is ``what is the
vector potential ${\vec A}_s$ coupled to the spin supercurrent ?''.
The answer to this question can be found in
the Aharonov-Casher (AC) effect \cite{AC} and
Dzyaloshinskii-Moriya (DM) interaction \cite{DM1,DM2}.
The conventional DM interaction\cite{DM2} is given by
\begin{equation}
H_{DM} = \sum_{<ij>} {\vec D_{ij}} \cdot ( {\vec S}_i \times {\vec S}_j ).
\end{equation}
When the DM vector ${\vec D}_{ij} = D_{ij} {\hat e}_z$, the total
Hamiltonian $H_{\rm total} = H_{XY} + H_{DM}$ with $H_{XY}$ in eq. (1) is
written as

\begin{equation}
H_{\rm total } = - \sum_{<ij>}
{{{\tilde J}_{\perp ij} } \over 2} ( e^{-i A_{ij}} S^+_i S^-_j +
e^{ i A_{ij}} S^+_i S^-_j)
\end{equation}
where ${\tilde J}_{\perp ij} e^{i A_{ij}} = J_{\perp ij}+ iD_{ij}$.
Therefore the DM vector ${\vec D}$ acts as the vector potential or
gauge field to the spin current. It is well known that the DM
interaction exists only when the inversion symmetry is broken at the
middle point between the two spins. Therefore when the crystal
structure has the inversion symmetry, the external electric field
${\vec E}$ induces the DM interaction. Namely ${\vec D}_{ij} \propto
{\vec E} \times {\vec e}_{ij}$, where ${\vec e}_{ij}$ is the unit
vector connecting the two sites $i$ and $j$. This form is identical
to the Aharanov-Casher (AC) effect, where the Lorentz transformation
of the electric field induces the magnetic field in the moving frame
which interacts with the spin moment. However the magnitude of the
coupling constant for AC effect is extremely small in vacuum since
it contains the rest mass of the electron $m c^2 \cong 5 \times 10^5
eV$ in the denominator. The situation is different for the DM
interaction in solids, i.e., the electrons are trapped in the strong
potential of the atoms with large momentum distribution leading to
the enhanced spin-orbit interaction. Therefore the gauge potential
$A_{ij}$ could be (a fraction) of the order of unity, e.g. $A_{ij}
\sim 2 \pi$ as seen below.

  To illustrate this,  consider the electron energy levels in the
ligand field of 3d-transition metal \cite{Tanabe}.
In the octahedral ligand field, the
d-orbitals are split into $e_g$ orbitals and $t_{2g}$ orbitals. The $t_{2g}$
orbitals, i.e., $d_{xy}$, $d_{yz}$, and $d_{zx}$, have energy lower than
$e_ g$ orbitals. If we take account of the spin degree of freedom, there is
6-fold degeneracy in $t_{2g}$ energy level. Due to the on-site spin-orbit
interaction, however, this degeneracy is lifted and we have two groups of
spin-orbit coupled states, labeled $\Gamma_7$ and $\Gamma_8$.
The 2-fold degenerate states, i.e., $\Gamma_7$, are given by

\begin{equation}
|a\rangle = \frac{1}{\sqrt3}(|d_{xy,\uparrow}\rangle+|d_{yz,\downarrow}\rangle+i|d_{zx,\downarrow}\rangle),
\end{equation}
and
\begin{equation}
|b\rangle= \frac{1}{\sqrt3}(|d_{xy,\downarrow}\rangle-|d_{yz,\uparrow}\rangle+i|d_{zx,\uparrow}\rangle),
\end{equation}
respectively, where the quantization axis of spin is taken to be the $z$
axis. For the sake of simplicity, we consider the above two states alone.
However, our method is valid for more general cases and one can easily
generalize it to any other spin-orbit strongly coupled situation.

We consider the case where the inversion symmetry exists at the 
middle point of the two magnetic ions, and the generic non-collinear magnetic 
ordering is realized by the competing exchange interactions $J$'s
and/or by the symmetry breaking due to the spin-orbit interaction.
Here the magnetic moment at $j$-th site points to the unit vector
${\vec e}_j = (\cos
\phi_j \sin \theta_j, \sin \phi_j \sin \theta_j, \cos \theta_j)$. The mean
field Hamiltonian applied to the Hubbard model is given by
( we take the unit where $\hbar =1$ hereafter):
$H = - U \sum_j {\vec e}_{j } \cdot {\vec S}_j$,
where $U$ is energy of Coulomb repulsion. For each site $j$, we restrict the
Hilbert space to the 2-dimensional one spanned  by the above two states,
and the effective Hamiltonian is reduced to the $ 2
\times 2$ matrix
\begin{equation}
-\frac{U}{3}
\left[
\begin{array}{cc}
-\cos\theta & \sin\theta e^{-i\phi}\\
\sin\theta e^{i\phi} & \cos\theta\\
\end{array}
\right].
\end{equation}
We diagonalize this Hamiltonian matrix to obtain
eigenstates $|P\rangle$, $|AP \rangle$ as
\begin{eqnarray}
|P\rangle &=&
\sin{\theta \over 2}|a\rangle +e^{i \phi} \cos {\theta \over 2} |b\rangle,
\nonumber \\
|AP\rangle &=&
\cos {\theta \over 2} |a\rangle - e^{i \phi}  \sin {\theta \over 2} | b\rangle.
\end{eqnarray}
Here $|P\rangle$ and $|AP \rangle$ means the spin state parallel
and anti-parallel to the unit vector ${\vec e}$, and the
corresponding eigenvalues are
$-\frac{U}{3}$ and $+\frac{U}{3}$, respectively.
For convenience, we define the coefficients $A^{i\sigma}$ and
$B^{i\sigma}$ and abbreviate the above two states as,
$|P\rangle = \sum_{i\sigma} A^{i\sigma}| d_{i\sigma}\rangle$,
$|AP\rangle = \sum_{i\sigma} B^{i\sigma}| d_{i\sigma}\rangle$,
where $i= xy,yz,zx, \sigma= \uparrow,\downarrow$.

\begin{figure}
  \begin{center}
    \includegraphics[width=8cm,clip]{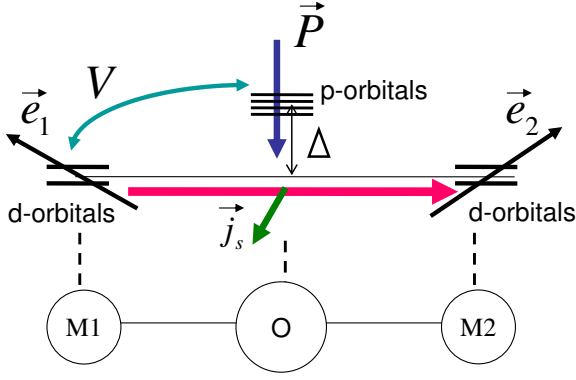}
    \caption{
The cluster model with two transition metal ions M1, M2 with the
oxygen atom O between them. With the non-collinear spin directions
${\vec e}_1$ and ${\vec e}_2$, there arises the spin current ${\vec
j}_s \propto {\vec e}_1 \times {\vec e}_2$ between M1 and M2. Here
the direction of the vector ${\vec j}_s$ (denoted by the green
arrow) is that of the spin polarization carried by the  spin
current. The direction of the electric polarization ${\vec P}$ is
given by ${\vec P} \propto {\vec e}_{12} \times {\vec j}_s$ where
${\vec e}_{12}$ is the unit vector connecting M1 and M2. }
  \end{center}
\end{figure}

 From now on, we focus on the three atom model as shown in Fig.1, which
represents the bond between the two transition metal ions M1, M2 through
the oxygen atom O. We take the hole picture below, where the
oxygen orbitals are empty.
We assume the generic case of  ${\vec e}_1$ and ${\vec
e}_2$ including the non-collinear configuration.
Each site has two states, i.e.,
$|P\rangle$ and $|AP\rangle$, mentioned above.
So we define $|P\rangle_j$
and $|AP\rangle_j $ $(j = 1,2)$
corresponding to the magnetic order on each site. Because of the existence
of the oxygen atom, there are hopping processes between the M site and the
O site. The transfer integrals between the d- and p-orbitals
can be found in the Slater-Koster tables\cite{Slater,Harrison},
and the hopping Hamiltonian is given as follows:
$$H_t=H_t^{1-m}+H_t^{m-1}+H_t^{2-m}+H_t^{m-2},$$

$$H_t^{1-m}=+V\sum_{\sigma}(p_{y,\sigma}^{\dagger}d_{xy,\sigma}^{(1)}+
p_{z,\sigma}^{\dagger}d_{zx,\sigma}^{(1)})=(H_t^{m-1})^{\dagger}$$
$$H_t^{2-m}=-V\sum_{\sigma}(p_{y,\sigma}^{\dagger}d_{xy,\sigma}^{(2)}+
p_{z,\sigma}^{\dagger}d_{zx,\sigma}^{(2)})=(H_t^{m-2})^{\dagger},$$
where $V(>0)$ is the transfer integral and the superscript $j$ denotes the
corresponding site number.
Let us now treat the above Hamiltonian perturbatively in $V$. The eight
bases we must prepare for this problem are $|P\rangle_j$,
$|AP\rangle_j , \ \ (j=1,2)$, and $p_{i,\sigma}, \ \ (i= y,z,\sigma=\uparrow,\downarrow)$.
Using the second-order perturbation theory, the four lowest lying states
and corresponding perturbed energies are obtained as follows:
$$|1\rangle = \frac{\alpha e^{-i\frac{\Delta \phi}{2}}}{\sqrt{2}|\alpha|}\biggl(|P\rangle_1
 +\frac{V}{\Delta}\sum_{\sigma}(A_{(1)}^{xy,\sigma}|p_{y,\sigma}\rangle+A_{(1)}^{zx,\sigma}|p_{z,\sigma}\rangle)\biggr)$$

$$
+\frac{1}{\sqrt{2}}\biggl(|P\rangle_2
-\frac{V}{\Delta}\sum_{\sigma}(A_{(2)}^{xy,\sigma}|p_{y,\sigma}\rangle+A_{(2)}^{zx,\sigma}|p_{z,\sigma}\rangle)\biggr),$$
with
$E_1=-\frac{4}{3}\frac{V^2}{\Delta}(1+|\alpha|),$
$$|2\rangle = -\frac{\alpha e^{-i\frac{\Delta \phi}{2}}}{\sqrt{2}|\alpha|}\biggl(|P\rangle_1
 +\frac{V}{\Delta}\sum_{\sigma}(A_{(1)}^{xy,\sigma}|p_{y,\sigma}\rangle+A_{(1)}^{zx,\sigma}|p_{z,\sigma}\rangle)\biggr)
$$
$$
+\frac{1}{\sqrt{2}}\biggl(|P\rangle_2
-\frac{V}{\Delta}\sum_{\sigma}(A_{(2)}^{xy,\sigma}|p_{y,\sigma}\rangle+A_{(2)}^{zx,\sigma}|p_{z,\sigma}\rangle)\biggr),$$
with
$E_2=-\frac{4}{3}\frac{V^2}{\Delta}(1-|\alpha|),$
and two other higher energy states.
Here $\Delta (>0)$ is the energy difference between the p-orbitals and
$|P\rangle_j$, $\Delta\phi = \phi_1-\phi_2$,
and we have introduced the complex number
$\alpha =
\cos{\frac{{\theta}_1}{2}}\cos{\frac{{\theta}_2}{2}}e^{-i\frac{\Delta \phi}{2}}
         +\sin{\frac{{\theta}_1}{2}}\sin{\frac{{\theta}_2}{2}}e^{+i\frac{\Delta \phi}{2}}$.
Before calculating the expected value of the polarization,
it is useful to note that only the
following matrix elements are non-zero from the shapes
of d- and p-orbitals;
$$
I =\int d^3{\vec r} d_{yz}^{(j)}({\vec r}) y p_z({\vec r}), \ \ (j=1,2),$$
and its cyclic permutations.
The integral $I$ is approximately estimated as $I\cong
\frac{16}{27}{Z_{\rm O}}^{5/2}{Z_{\rm M}}^{7/2}(\frac{Z_{\rm
O}}{2}+\frac{Z_{\rm M}}{3})^{-6}a_0$ , where $a_0$ is Bohr
radius and $Z_{\rm O}/Z_{\rm M}$ is the atomic number of O/M.
We can easily check the above results by expanding wavefunctions in terms
of lattice constant $a$. So let us now calculate the expected value of
polarization in the following two cases.

{\it Double-exchange interaction}\cite{Anderson1,kanamori}.---
First, we consider the situation where only one hole is present. In
this case, this hole is put into the ground state, determined by the
above second order perturbation theory, and the expected value of
polarization, $\langle1 | er |1 \rangle/\langle1|1\rangle$, is given
by
\begin{equation}
{\vec P} \cong  -{{e V} \over {3 \Delta}} I
 {{{\vec e}_{12} \times   { ({\vec e}_1 \times {\vec e}_2 )}}
\over { | \cos { {\theta_{12}} \over 2} |}}
\end{equation}
where ${\vec e}_{12}$  is the unit vector parallel to the direction
of the bond from site M1 to site M2, and $\theta_{12}$ is the angle
between the two vectors ${\vec e}_1$ and ${\vec e}_2$, i.e., ${\vec
e}_1 \cdot {\vec e}_2 = \cos \theta_{12}$. Spin current ${\vec J}_s$
is approximately given by ${\vec J}_s \sim (V^2/\Delta)a_0 ({\vec
e_1} \times {\vec e}_2)/\cos(\theta_{12}/2)$, and eq.(10) can be
rewritten as ${\vec P} \sim (e/V) {\vec e}_{12} \times {\vec j}_s$.
Therefore the spin current is essential to the electric
polarization.

{\it Superexchange interaction}\cite{Anderson2}.---
Next we consider the case of two holes. From a
viewpoint of Hartree-Fock approximation, two holes are put into the
ground state $|1\rangle$ and the second low-lying state $|2\rangle$,
and the expected value of the polarization is given by the following
form similar to (10):
\begin{equation}
{\vec P} \cong  -{ {4e} \over 9} {( {V \over \Delta})}^3 I {\vec e}_{12}
\times( {\vec e}_1 \times {\vec e}_2 ).
\end{equation}
In this case, we must pay attention to the difference of the
normalization factor between the two perturbed states, i.e,
${\langle 1|1 \rangle}^{-1} \cong 1-\frac{2}{3}
({\frac{V}{\Delta}})^2(1-|\alpha|)$ and ${\langle 2|2 \rangle}^{-1}
\cong 1-\frac{2}{3} ({\frac{V}{\Delta}})^2(1+|\alpha|)$. Of course
there are many other terms but this term mainly contributes. As the
above result is order of $(V/\Delta)^3$, one may think that
the polarization is too small to observe. By using the superexchange
interaction J, i.e., $J \cong V^4/(U\Delta^2)$, however, (11) can be
rewritten as
\begin{equation}
{\vec P} \cong  -{ {4e} \over 9} {J \over V}{ U \over \Delta} I {\vec e}_{12} \times( {\vec e}_1 \times
{\vec e}_2 ).
\end{equation}
Again this equation can be interpreted in temrs of the spin current
${\vec j}_s$ as
${\vec P}\sim (eU/V \Delta) {\vec e}_{12} \times {\vec j}_s$.
Therefore the magnitude is not too small on the ground that
the ratio of $U$ to $\Delta$ is by no means small practically.

For more general models, the magnitude of the electric
polarization induced by the spin current would differ from
those obtained in the present model, and depends on the
details of the electronic level structure. Most likely
it will be smaller than in the present case, especially
for the $e_g$ systems where the spin-orbit interaction
is smaller. However the geometrical relation between
the spin current and electric polarization remains
unchanged.


\begin{figure}
 \begin{center}
  \includegraphics[width=8cm,clip]{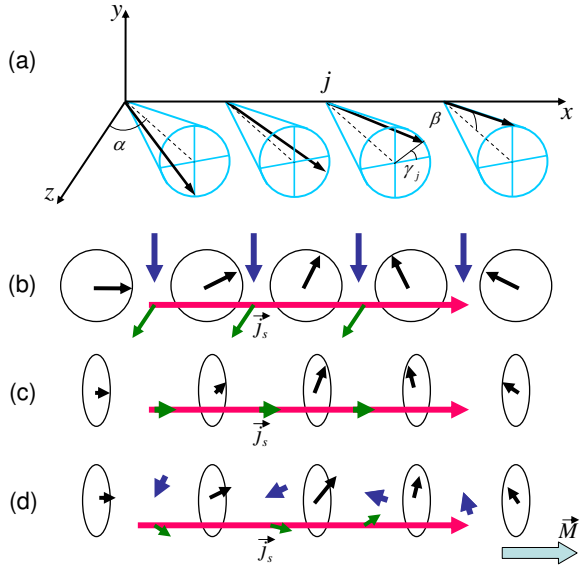}
  \caption{(a) Generic spiral spin configuration. (b-d) Some of the specific configurations
where the geometrical relation among spins(black arrows), spin current(green arrows), and electric polarization
(blue arrows) are shown.
}
 \end{center}
\end{figure}

{\it Applications to the spiral magnets}--- Now we turn to the
discussion on realistic non-collinear magnets. One of the typical
examples is the spiral structure where the direction of the spin
rotates along the wavevector ${\vec q}$. Figure 2(a) shows the most
generic spiral spin configuration where the spiral axis is along the
$x$-axis, the cone axis direction have the angle $\alpha$ measured
from the $z$-axis, and the cone angle is $\beta$. The angle
$\gamma_j$ for the $j$-th spin direction is measured from the
$zx$-plane, and the undistorted spiral means $\gamma_j = q j +
\gamma_0$. The spin at site $j$ can be written as ${\vec S}_j = S(
\cos \beta \sin \alpha + \sin \beta \cos \alpha \cos \gamma_j, \sin
\beta \sin \gamma_j, \cos \beta \cos \alpha - \sin \beta \sin \alpha
\cos \gamma_j)$. With this configuration one can easily calculate
the spin current ${\vec j}_{s j+1/2}$ and resultant electric
polarization ${\vec P}_{j+1/2}$ at each link connecting $j$ and
$j+1$ as
$ (P_{j + 1/2})_x = 0$,
$ (P_{j + 1/2})_y \sim
- \cos \beta \sin \beta \sin \alpha
[ \sin \gamma_{j+1} - \sin \gamma_j ] - \sin^2 \beta \cos
\alpha \sin ( \gamma_{j+1} - \gamma_j)$,
$ (P_{j + 1/2})_z \sim \cos \beta \sin \beta
[ \cos \gamma_{j+1} - \cos \gamma_j ] $.
Therefore only the $y$-component
of the uniform electric polarization ${\vec P} = \sum_j {\vec P}_{j+1/2}$
is nonzero and is given by
$P_y \propto   \sum_j \sin^2 \beta \cos
\alpha \sin ( \gamma_{j+1} - \gamma_j)$.

Figure 2(b-d) shows some typical cases and their spin current and electric
polarization. When the spins are within the $xy$-plane as in Fig. 2(b),
i.e., $\alpha=0$, $\beta=\pi/2$,
the spin current ${\vec j}_s$ is along the
$z$-direction, and the electric polarization ${\vec P}$
is along the $y$-direction
for each site. Therefore the total uniform polarization is finite
along the $y$-direction. Note that even when the spiral wavenumber
${\vec q}$ is incommensurate with the lattice periodicity, the uniform
polarization, i.e., the ferroelectricity, is realized.
When the spins are in the $yz$-plane (Fig. 2(c))
, i.e., $\alpha = \pi/2$, $\beta= \pi/2$,
both ${\vec e}_{j j+1}$ and ${\vec j}_s$ are in the
$x$-direction and their vector product is zero. Therefore we do not expect
any electric polarization induced in this case.
Figure 2(d) shows the ``conical''spin structure, where the finite
$x$-component of the spin is induced starting from the structure in
Fig. 2(c), i.e., $\alpha = \pi/2$, $0<\beta<\pi/2$.
In this case, the finite $S_x$ component
at each site produces the rotating polarization, but these cancel
out to zero uniform electric polarization.

Now consider the effect of the external electric field $E_y$ along
the $y$-direction on  the generic spiral configuration. This field
induces  uniform magnetization (per site),  estimated as ${\vec m}_x
\propto E_y \sin^2 \beta \cos \beta \sin \alpha \cos \alpha \sin q$,
and ${\vec m}_z \propto E_y \sin^2 \beta \cos \beta \sin q ( a + b
\cos^2 \alpha)$ ($a,b$: constants). This dependence is consistent
with the group theoretical consideration and the experiments on
ZnCr$_2$Se$_4$ ( Figures 6 and 7 of ref. \cite{shiratori} ).

{\it Acknowledgements} This work was supported by NAREGI and Grant-in-Aids for Scientific Research from MEXT, and US DOE
LDRD at Los Alamos.

\end{document}